\documentstyle[aps,preprint]{revtex}
\begin{document}
\draft
\preprint{ISSP Sep. 22, 1994}
\title{
Universal Correlations in the random matrices and 1D particles with long range
interactions
in a confinement potential}
\author{Y. Morita$^{*}$, Y. Hatsugai$^{**}$ and M. Kohmoto}
\date {September 22, 1994}
\address{
 Institute for Solid State Physics,
 University of Tokyo,
 7-22-1 Roppongi Minato-ku, Tokyo 106, Japan
}
\maketitle
\begin{abstract}
We study the correlations between eigenvalues of the large random matrices by a
renormalization group
approach. The results strongly support the universality of the correlations
proposed by Br\'ezin and Zee.
Then we apply the results to the ground state of the 1D particles with long
range interactions
in a confinement potential.
We obtain the exact ground state.
We also show the existence of a transition similar to a phase separation.
Before and after the transition, we obtain
the density-density correlation explicitly. The correlation shows nontrivial
universal behavior.
\end{abstract}
\pacs{}
\narrowtext

The random matrix theory has a long and distinguished
history $\cite{Metha}$ and the results have been applied to many fields,
including heavy nuclei, quantum gravity and mesoscopic systems
$\cite{gravity}$ $\cite{Lee}$.

Recently Br\'ezin and Zee $\cite{B&Z1}$ $\cite{B&Z2}$ have proposed the
universal correlations between eigenvalues for a broad class of
large random matrices by calculating them explicitly with an ansatz for the
orthogonal polynomials.
What is remarkable in their results is that, except for an overall scale
setting, the width of the correlations is completely independent of
the details of the Hamiltonians. Beenakker $\cite{Bee}$ extended their results
by a different technique.
As is suggested by Br\'ezin and Zee, we try to understand this through a
renormalization group (RG) approach.
Following the RG method developed by Br\'ezin and Zinn-Justin $\cite{RG}$, we
calculate the $\beta$ function for a broad class of
unitary ensembles and show the existence of the stable Gaussian fixed point.
This result strongly supports the existence of the universality
in the large random matrices proposed by Br\'ezin and Zee.

 Moreover we consider the 1D particles with long range interactions in a
confinement
potential using the random matrix theory. The exact ground state is obtained.
We also obtain the density distribution and the density-density correlation
explicitly and investigate them.
The ground state shows a novel transition to the ``separation'' phase as
described below.

Let us consider the random matrix model described by the Hamiltonian
\begin{equation}
H=\frac {1}{2}\mathop{\rm{Tr}} M^2 +
\sum_{K=2}^{\infty }[\frac {p_K}{N^{2K-2}}(\mathop{\rm{Tr}} M^2)^K+\frac
{q_K}{N^{K-1}}(\mathop{\rm{Tr}} M^{2K})],
\end{equation}
where $M$ is an $N$ by $N$ Hermitan matrix.
Set one of $p_K$'s or $q_K$'s to be $g$( $\ne 0$ ) and the rest to be zero.
The free energy of this model is $ F(g)=-\lim_{N \rightarrow
\infty}\frac{1}{N^2}\ln [\int dM \exp(-H)] $,
where $dM$ is ${\cal N}\prod_idM_{ii}\prod_{i<j}dReM_{ij} dImM_{ij}$ and ${\cal
N}$ is set to satisfy $F(g=0)=\frac {1}{2}$.
Then we have the differential equation $\cite {S.Hikami}$ $\cite {RG}$
\begin{equation}
(\sum_{K=0}^{\infty}a_Kg^K)(\frac{\partial F}{\partial
g})=-g(\sum_{K=0}^{\infty}b_Kg^K)(\frac{\partial^2F}{\partial g^2})+
(\sum_{K=0}^{\infty}c_Kg^K),
\end{equation}
and the $\beta$ function is given by $\cite {S.Hikami}$
\begin{equation}
\beta(g)=-g\frac{\sum_{K=0}^{\infty}b_Kg^K}{\sum_{K=0}^{\infty}a_Kg^K.}
\end{equation}
Equation (2) with (3) denotes the large-$N$ limit of the Callan-Symanzik-like
equation. The idea is an analogy
between the matrix size in the matrix model theory and the momentum cut off in
the quantum field theory, which the double scaling limit in
the 2D quantum gravity suggests.
The $\beta $ function has two zero points, one is the Gaussian fixed point
($g=0$) and the other is the fixed
point ($g<0$) which governs the double scaling limit in the 2D quantum gravity
$\cite {RG}$. The latter fixed point, however, does not play
any role in the present problem since the coupling constant $g$ is negative.

The crucial step for understanding the universal correlations is to investigate
the stability of the Gaussian fixed point.
In the following we put $a_0=1$ without loss of generality.
Hikami $\cite {S.Hikami} $ investigated the case when $q_2$ is $ g $ and
obtained
\begin {equation}
\beta (g)=-\frac {g(1+48g)}{3(1+24g).}
\end {equation}
This is consistent with the known exact results obtained by different
techniques $\cite{BIPZ}$.
For example, (4) implies that a phase transition takes place at $g=-\frac
{1}{48}$, while the exact critical value of $g$
is also $-\frac {1}{48} $.

If we have the complete series expansion of $F(g)$, it is possible to calculate
the $\beta $ function in principle.
In general, however, it is technically difficult to obtain high order terms.
Thus we shall calculate the series expansion up to low order and determine the
values of $a_1, b_0
$ and $b_1$. We have set $a_K=b_K=c_{K-1}=0$ for $K=2,3,\cdots$ .
They are still expected to be good approximations, as demonstrated in several
cases$\cite{S.Hikami}$.

We consider two classes of models.

i) $ H=\frac{1}{2}\mathop{\rm{Tr}} M^2+\frac{g}{N^{2n-2}}(\mathop{\rm{Tr}}
M^2)^n $.

At first we calculate the series expansions of $F(g)$. $F(g)$ is
\begin {equation}
 F=-\lim_ {N\rightarrow \infty}\frac {1}{N^2}\ln [\int dx\, c\,
x^{\frac{N^{2}}{2}}\exp (-\frac {1}{2}x-\frac{g}{N^{2n-2}}x^n)],
\end {equation}
where $x$=$Tr M^2$ and $c$ is a constant.
By replacing $x$ by $N^2x$, we obtain the free energy by the saddle point
method as
$F=\frac {1}{2}x +gx^n -\frac {1}{2} \ln x$.
The saddle point equation is
$\frac{1}{2}x+ngx^n-\frac {1}{2}=0$.
Thus we have $\frac {\partial F}{\partial g}=1-2n^2g+O(g^2)$ in the small $g$
limit and $ \frac {\partial F}{\partial g}=\frac {1}{2ng}-
\frac {1}{(2n)^{\frac {n+1}{n}} g^{\frac {n+1}{n}}}+O(\frac {1}{g^{\frac
{n+2}{n}}})$ in the large $g$ limit.

Putting these series expansions into (2), we obtain the $\beta $ function from
(3) as
\begin{equation}
\beta (g)=-\frac{g(1+2n^3g)}{n[1+(2n^2+2n)g].}
\end{equation}
When $n$=2, this is the exact solution of (2) and (3) and consistent with the
exact results obtained by other techniques $\cite{vector}$.
Thus we expect that the $\beta $ function for $n$=2 is exact.

ii) $ H=\frac{1}{2}\mathop{\rm{Tr}} M^2+\frac{g}{N^{n-1}}\mathop{\rm{Tr}}
M^{2n} $.

Since the matrix $M$ can be diagonalized by a unitary matrix, the free energy
can be expressed by the eigenvalues. Thus one can have
the complete series expansion of $F(g)$ for $n$=2 $\cite {BIPZ}$ and the $\beta
$ function is obtained as (4). It is, however, difficult
to get the exact results for higher $n$.
Instead we exactly map the Hamiltonian to $H=\frac{1}{2}\mathop{\rm{Tr}}
M^2+\sum_{K=1}^\infty \frac{d_Kg^K}{N^{2nK-2}}(\mathop{\rm{Tr}} M^2)^{nK}$,
integrating out
the angular part of the Hamiltonian $\cite {S.Hikami}$. In the lowest order in
$g$, the Hamiltonian is approximated by
$H=\frac{1}{2}\mathop{\rm{Tr}} M^2+\frac{d_1g}{N^{2n-2}}(\mathop{\rm{Tr}}
M^2)^{n}$,
where $d_1$ is obtained by comparing $F$ before and after the mapping.
The result is $d_1=2(n=2), 5(n=3), 14(n=4), 42(n=5), 132(n=6) ,\cdots$.
By following the same procedure as (i), we obtain
\begin{equation}
\beta (g)=-\frac{g(1+2n^3d_1g)}{n[1+(2n^2+2n)d_1g].}
\end{equation}

The results (6) and (7) show that the Gaussian fixed point ($g$=0 ) is stable
for a broad class of large random matrices ( see Fig. 1 ).
This simple renormalization group argument leads to a direct and intuitive
understanding of the universal properties of
the large random matrices $\cite{B&Z1}$ $\cite{B&Z2}$.

Next we shall apply the results obtained above to 1D $N$-body Fermi or Bose
systems in a confinement potential.
The Hamiltonian is
\begin{equation}
H=H_{I}+H_{II}+H _{III},
\end{equation}
\begin{equation}
H_{I} =
-\sum_{i=1}^{N}\frac{d^{2}}{dx_{i}^2}+\sum_{i<j}\frac{g_0}{(x_{i}-x_{j})^2},
\end{equation}
\begin{equation}
H_{II}=\sum_{i} (Ax_{i}^2+Bx_{i}^4+Cx_{i}^6),
\end{equation}
and
\begin{equation}
H_{III}=\sum_{i<j}D(x_{i}-x_{j})^2,
\end{equation}
where $H_{I}$ is the Hamiltonian of the $N$-body system with $\frac{1}{r^2}$
interaction, $H_{II}$ is a confinement potential and $H_{III}$
is attractive harmonic interaction.
Note that $H_{II}$ is double-well like when B is negative and C is positive.
Let us parameterize $A, B, C, D$ and $ g_0 $ by
\begin{eqnarray}
A&=&\omega^2-12\alpha-12(N-1)\alpha\lambda , \\ B&=&8\alpha \omega ,\\
C&=&16\alpha^2 , \\ D&=&4\alpha \lambda,
 \\ g_0&=&2\lambda ^2-2\lambda ,
\end {eqnarray}
 where we set $\alpha $ to be positive.
Although the parameters $A,B,C,D$ and $g_0$ are not totally independent, we can
change the depth of the double well
without changing the interaction by changing $\omega $, while keeping $\alpha $
and
$\lambda $ fixed, where $\omega $ can also be negative and the new universality
class described below appears.
At $\alpha =0 $ , we have the Calogero-Moser-Sutherland (CMS) model
$\cite{Sutherland}$ $\cite{Calogero}$ $\cite{Moser}$
, which is known to be integrable and has been studied extensively. The system
described by $H=H_I+H_{III}$ was also described
by Calogero$\cite{Calogero}$.
Thus we call (8) with (9), (10) and (11) the generalized
Calogero-Moser-Sutherland (GCMS) model.
We can also generalize $H_{II}$ and $H_{III}$ in a way that the extended
Hamiltonian still has the universal properties investigated below,
but we shall restrict our attention to (8) for simplicity.

It can be shown that the following Jastrow-type function is an exact
eigenfunction with $ E=\omega N [1+\lambda (N-1)]$
\begin{equation}
\Psi(x_{1},\cdots ,x_{N})=S\prod_{i<j}|x_{i}-x_{j}|^{\lambda}\exp
(-\frac{1}{2}\omega\sum_{i=1}^{N}x_i^2-\alpha\sum_{i=1}^Nx_i^4).
\end{equation}
The factor S is equal to 1 for bosons, while for fermions $S=(-1)^P$ is the
parity of the particle ordering
permutation $P$, $x_{P_1}<x_{P_2}<\cdots<x_{P_N}$.
It is the ground state  since, for each ordering of the particles, this
wavefunction is nodeless
$\cite{Sutherland}$ $\cite{Calogero}$. Thus we obtain the exact ground state.
Note that even for the much simpler system of non-interacting particles in an
anharmonic potential,
the ground state can not be obtained exactly.
The squared amplitude of the wavefunction is
\begin{equation}
|\Psi(x_{1},\cdots ,x_{N})|^2=\prod_{i<j}|x_{i}-x_{j}|^{\beta}\exp
(-\omega\sum_{i=1}^{N}x_i^2- 2\alpha\sum_{i=1}^Nx_i^4),
\end{equation}
where  $\beta$ is 2$\lambda$.

For $\beta $=1, 2 and 4, (18) is identical to the probability weight of the
orthogonal, unitary and symplectic ensembles
in the random matrix theory respectively, where we identify the eigenvalues of
random matrices with the particle coordinates.
Thus the following arguments are also applicable to the random matrix theory.

The exact ground state (18) has nontrivial structures.
In order to investigate them we take the limit $ N\rightarrow \infty $, $
\frac{\omega}{N}\rightarrow \omega_0 \sim O(1)$ and
$\frac{\alpha}{N} \rightarrow \alpha_0\ \sim O(1)$,
taking into account the effect of the confinement potential.
At first, we consider the density distribution $\rho (x)$.
In the large-$N$ limit we can apply the saddle point method to (18) and obtain
\begin{equation}
\omega_0 x+4\alpha_0 x^3=\frac{\beta}{2N}P\int dy \frac{\rho(y)}{x-y ,}\
\end{equation}
where $P$ denotes the principal part of the integral.
This integral equation can be sloved by the Hilbert transformation
$\cite{BIPZ}$ and we have
(see Fig. 2),

i) for $ \omega_0  > \omega_c $
\begin{equation}
 \rho (x ) = \left\{
                \begin{array}{@{\, }ll}

\frac{N}{\pi}(\frac{8\alpha_0}{\beta}x^2+\frac{2\omega_0}{\beta}+
\frac{4\alpha_0 a^2}{\beta})(a^2-x^2)^\frac{1}{2}
& (-a<x<a )\\
                    0 & \mbox{(otherwise)}\\
                \end{array}
                   \right.
\end{equation}

ii) for $ \omega_0  <  \omega_c $
\begin{equation}
\rho (x) = \left\{
               \begin{array}{@{\, }ll}
                   \frac{N}{\pi}\frac{8\alpha_0}{\beta }|x
|[(d^2-x^2)(x^2-c^2)]^\frac{1}{2} &  (c<|x|<d) \\
                    0 & \mbox{(otherwise)},\\
               \end{array}
                  \right.
\end{equation}
with $ a^2 =
\frac{\beta}{24\alpha_0}[-\frac{4\omega_0}{\beta}+(\frac{16\omega_0^2}{\beta^2}
+\frac{192\alpha_0}{\beta})^{\frac{1}{2}}],
c^2=-\frac{\omega_0}{4\alpha_0}-\sqrt{\frac{\beta}{4\alpha_0}},
d^2=-\frac{\omega_0}{4\alpha_0}+\sqrt{\frac{\beta}{4\alpha_0}}$ and
$ \omega_c=-\sqrt{4\alpha_0\beta} $.

When $\omega_0 < \omega_c$, the paticles become separated (see Fig.2). At
$\omega_c$, the universality class of the system changes.
We call this novel phenomenon ``separation''. A similar phenomenon is the
phase separation in the $t$-$J$ model $\cite {Sepa}$.

Secondly, we consider the density-density correlation of the GCMS model.
As a direct consequence of the Br\'ezin and Zee's results $\cite {B&Z1}$ $\cite
{B&Z2}$ and the extended results by Beenakker $\cite {Bee}$,
we have
\begin{eqnarray}
\rho_c(x, y) = -\frac{1}{\beta \pi^2a^2}f\left ( {x\over a}, {y\over a}\right )
,\\
f(x, y) = \frac{1}{(x-y)^2} \frac{(1-xy)}{\sqrt{(1-x^2)(1-y^2) ,}}
  \label{la.univ}
\end{eqnarray}
where $|x - y|\sim O(1)$ and  $\rho_c(x, y)$ is defined by
$<\sum_i\delta (x-x_i)\sum_j\delta (y-x_j)>-<\sum_i\delta (x-x_i)><\sum_j\delta
(y-x_j)>$.
Notice that there is the effect of the confinement potential in (22),
since (22) is not the function of only $(x-y)$.
The importance of this result is that $\rho_c(x, y)$ is universal in contrast
to $\rho (x)$.
Although (22) is correct for $\omega _0 > \omega _c$, the numerical simulation
$\cite {Koba}$ suggests that it does not hold for
$\omega _0 < \omega _c$, where the potential has a deep double well. We expect
that the system belongs to a different universality class for
$\omega _0 < \omega _c$.
In order to show this explicitly, we calculate $\rho_c(x, y)$ after separation
using the method developed recently by Beenakker$\cite{Bee}$.
Let us consider the following wavefunction in which the support of $\rho$($x$)
is $(p, q)\cup (r, s)$ ($p<q<r<s$) (see Fig. 3)
\begin{equation}
|\Psi (x_1,\cdots , x_N)|^2=\exp[\beta\sum_{i<j}\ln |x_i-x_j|-\beta
\sum_iV(x_i)] ,
\end{equation}
where $V(x )$ is a function.
In the large-$N$ limit we can apply the saddle point method to (24) and obtain
\begin{equation}
P\int_{p}^{q}dy \frac{\rho (y)}{x -y}+P\int_{r}^{s}dy \frac{\rho
(y)}{x-y}=\frac{dV(x) }{dx}.
\end{equation}
A variation of (25) gives
$ P\int_{p}^{q}dy \frac{\delta \rho (y)}{x-y}+P\int_{r}^{s}dy \frac{\delta \rho
(y)}{x-y}=\frac{d\delta V(x) }{dx}. $
Since separation does take place and there are no particles in the finite range
between $q$ and $r$, we have
\begin {equation}
 \delta \int_{p}^{q}dx\rho (x)=\delta \int_{r}^{s}dx\rho (x)=0.
\end {equation}
Thus, from the fact that $\rho _c(x, y)$ is symmetric under an exchange of $x$
and $y$, and  the Beenakker formula $\cite{Bee}$
$\rho _c(x,y)=-\frac{1}{\beta }\frac{\delta \rho (x)}{\delta  V(y) ,}$
we have
\begin{equation}
     \rho_c(x, y) = \left\{
                                      \begin{array}{@{\, }ll}
                                        A_0\frac{1}{(x-y)^2}
\frac{\frac{1}{2}(p+q)(x+y)-pq-xy}{\sqrt {(x-p)(q-x)(y-p)(q-y)}} & ( x,y\in [p,
q] )  \\
                                        B_0\frac{1}{(x-y)^2}
\frac{\frac{1}{2}(r+s)(x+y)-rs-xy}{\sqrt {(x-r)(s-x)(y-r)(s-y)} } & ( x,y\in
[r,s] )\\
                                        0 & \mbox{(otherwise),}
                                      \end{array}
                                         \right.
\end{equation}
where $|x-y|\sim O(1)$ and $A_0$ and $B_0$ are constants. Note that $\rho _c(x,
y)$ is still universal after separation.

In summary, we calculated the $\beta$ functions for a broad class of random
matrices and found that the Gaussian fixed point is stable .
As is suggested by Br\'ezin and Zee$\cite {B&Z1}$ $\cite {B&Z2}$, the stable
Gaussian fixed point is responsible for
the universality of the correlations in the large random matrices.
This result is interpretated in the GCMS model (8). We investigated the system
in the large $N$ limit. The system can be
in the two universality classes depending on the parameters of the Hamiltonian.
We call this change of the universality class ``separation''.
This phenomenon is similar to the phase separation in the $t$-$J$ model. In all
the parameter regions , we obtained the density distribution
and the density-density correlation explicitly. The density-density correlation
shows the universal behavior
in contrast to the density distribution. We take into account the effect of the
confinement potential and
our results may have some implications
to the mesoscopic systems such as quantum wire.

We wish to thank S. Hikami, T. S. Kobayakawa, and A.Zee for very helpful
discussions.

\begin{figure}
Fig. 1. The $\beta $ function of the random matrix model (schematic).

The arrow denotes the renormalization group flow .
There are two fixed points.
One is the stable Gaussian fixed point and the other is the unstable fixed
point
which governs the double scaling limit in the 2D gravity $\cite {gravity}$ .

Fig. 2. The density distribution of the generalized Calogero-Moser-Sutherland
model (schematic).

$x$ denotes the space coordinate and $\rho (x)$ the density distribution.

Fig. 3.  The density distribution of the wavefunction (24) (schematic).

$x$ denotes the space coordinate and $\rho (x)$ the density distribution.

\end{figure}


\begin{references}

\bibitem[*] {ko} electronic address : {\tt morita@kodama.issp.u-tokyo.ac.jp}

\bibitem[**]{hmoto} electronic address : {\tt hatsugai@tansei.cc.u-tokyo.ac.jp}

\bibitem{Metha} M. L. Metha, {\it Random Matrices}, 2nd ed.  (Academic, New
York, 1991)

\bibitem{gravity} F. David, {\it Nucl. Phys. B} {\bf 257}, 45 (1985)

\bibitem{Lee} B. D. Simons, P. A. Lee, and B. L. Altschuler, {\it Phys. Rev.
Lett.} {\bf 70}, 4122, (1993)

\bibitem{B&Z1} E. Br\'ezin and A. Zee, {\it Nucl. Phys. }{\bf 402(FS)}, 613
(1993)

\bibitem{B&Z2} E. Br\'ezin and A. Zee, {\it Compt. Rend. Acad. Sci. }{\bf 317},
735 (1993)

\bibitem{Bee} C. W. J. Beenakker, preprint cond-mat/9310010

\bibitem{RG} E. Br\'ezin and J. Zinn-Justin, {\it Phys. Lett. B }{\bf 288}, 54
(1992)

\bibitem{S.Hikami} S. Hikami, preprint UT-KOMABA/94/13 June 1994

\bibitem{BIPZ} E. Br\'ezin, C. Itzykson, G. Parisi and J. B. Zuber, {\it Comm.
Math. Phys. }{\bf 59}, 35 (1978)

\bibitem{vector} S. Nishigaki and T. Yoneya, {\it Nucl. Phys. B }{\bf 348}, 787
(1991)

\bibitem{Sutherland} B. Sutherland, {\it J. Math. Phys. }{\bf 12}, 246 (1971);
{\it J. Math. Phys. }{\bf 12}, 251 (1971)

\bibitem{Calogero}  F.Calogero, {\it J. Math. Phys.} {\bf 10}, 2197 (1969)

\bibitem{Moser} J.Moser, {\it Adv. Math.} {\bf 16}, 197 (1975).

\bibitem{Sepa} V. J. Emery, S. A. Kivelson, and H. Q. Lin, {\it Phys. Rev.
Lett. } {\bf 64}, 475 (1990)

\bibitem{Koba} T. S. Kobayakawa, Y. Hatsugai, M. Kohmoto and A. Zee, preprint






























\end{references}
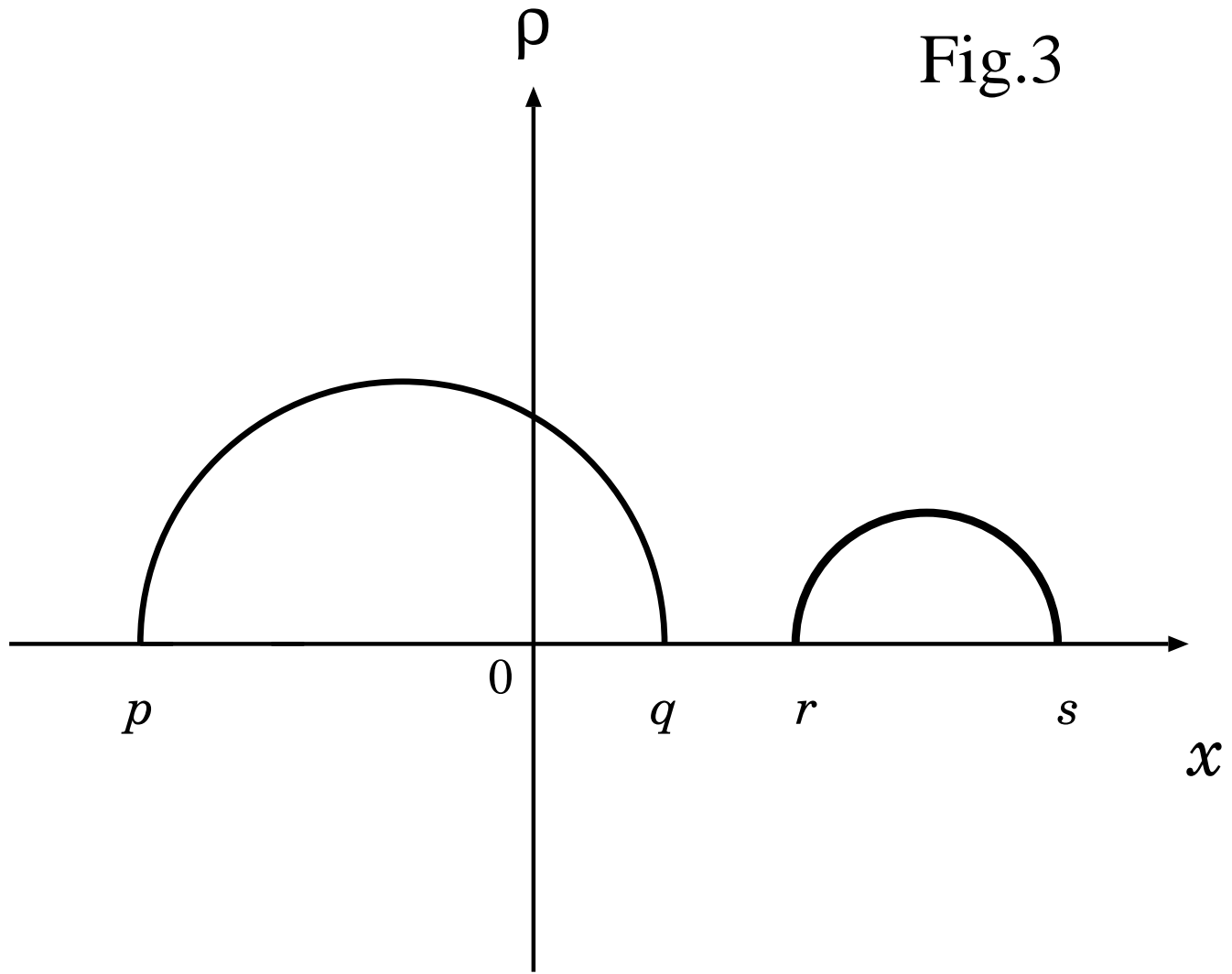
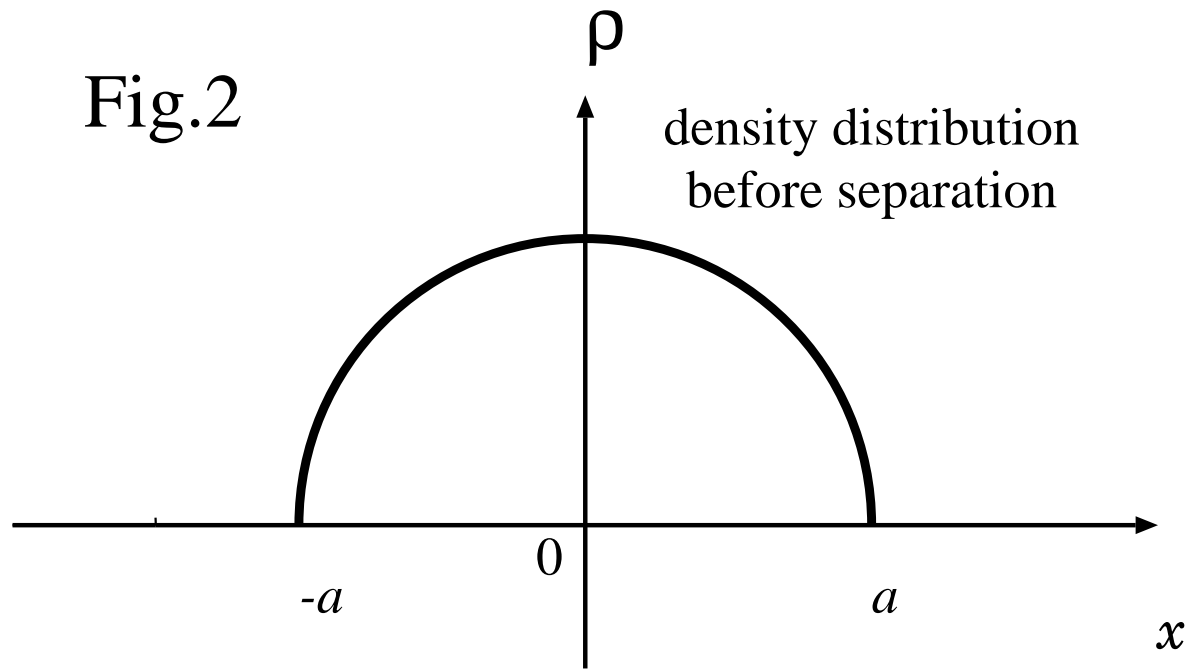
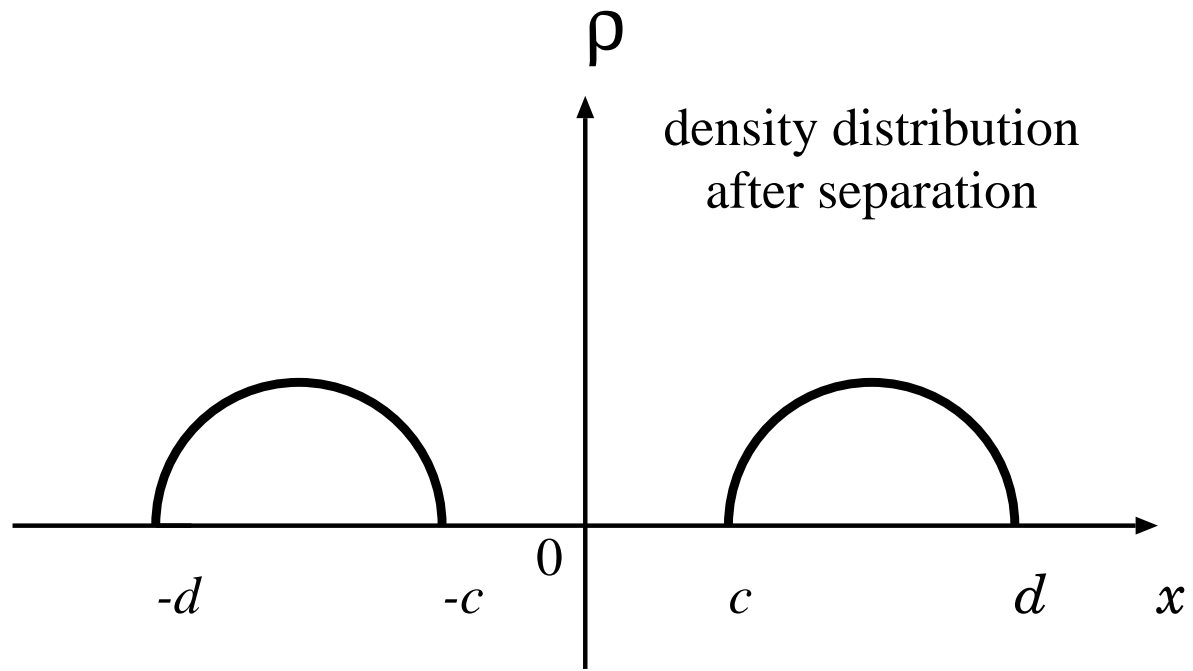
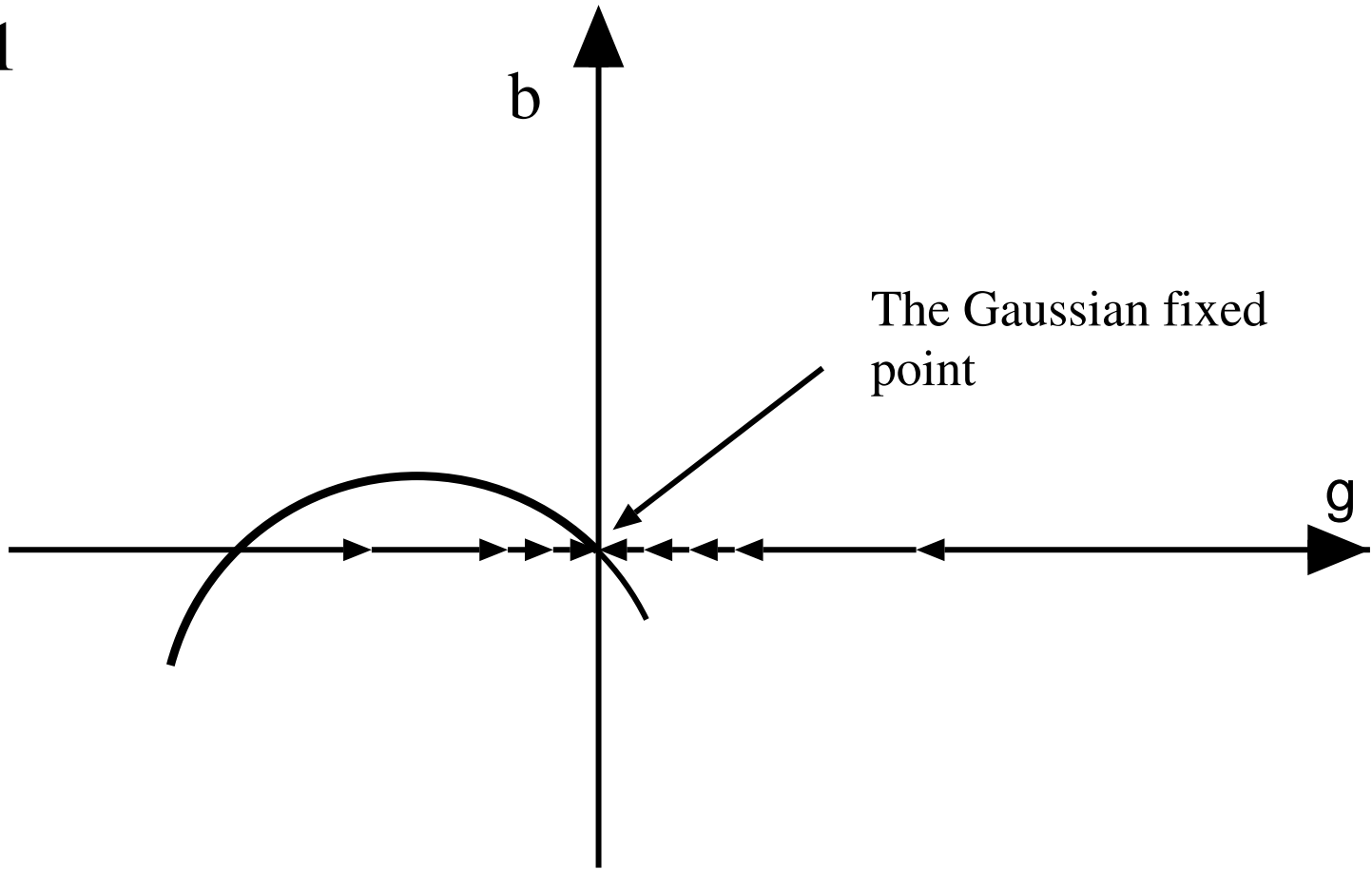
\end{document}